%% file: CH4-DM.tex
\documentclass[aps,prd,12pt]{revtex4}
\begin{document}
%%%%%%%%%%%%%%%%
\def\erf{\mbox{erf}}
\def\ER{E_{\rm R}}
\def\SD{{\rm SD}}
\def\SI{{\rm SI}}
\renewcommand{\baselinestretch}{1.0}
%%%%%%%%%%%%%%%%%%%%%%%%%%%%%%%%%%%%%%%%%%%%%%%%%%%%%%%%%%%%%%%%%
\title{Methane ionization chamber to search for 
        spin-dependent dark matter interactions} 
\author{B.M. Ovchinnikov and V.V. Parusov}
\affiliation{Institute for Nuclear Research of RAS, 
            %60-th Oct. Anniversary prospect 7a, 
            Moscow 117312, Russia}
\author{V.A.~Bednyakov}
\affiliation{Dzhelepov Laboratory of Nuclear Problems,
             Joint Institute for Nuclear Research, %\\
             141980 Dubna, Russia} 

\begin{abstract} %%%%%%%%%%%%%%%%%%%%%%%%%%%%%%%%%%%%%%%%%%%%%%%%%
       A liquid-methane ionization chamber is proposed
       as a setup to search for spin-dependent interactions of 
       dark-matter particles with hydrogen.
\end{abstract} %\pacs{11.30.Pb,12.60.Jv,95.35.+d} %%%%%%%%%%%%%%%%

\maketitle 

       Search for relic dark-matter particles (DM) and 
       investigations of their properties are among the 
       most challenging problems of modern astrophysics, cosmology
       and particle physics. 
       Detection of DM particles will cast %spread
       light on the structure and genesis of our Universe.   
       Therefore a lot of experiments (see for example 
\cite{Bernabei:2003za,Benoit:2002hf,Morales:2002ud,%
Baudis:1999aa,Klapdor-Kleingrothaus:2003pe}) 
       are aimed at detection of these DM particles today.
       In modern particle physics 
       the lightest supersymmetric particle (LSP) neutralino is 
       assumed to be one of the best DM candidate.  
       The promising idea of LSP detection relies on
       rather weak, but not vanishing, spin-dependent (SD) and 
       spin-independent (SI) interactions of the LSP with ordinary
       matter nucleus. 
       In general, one believes 
       that for heavy target nuclei the SI interaction
       dominates, but for light nuclei the SD interaction   
       makes the dominant contribution 
       to the expected event rate of DM detection
(see for example 
\cite{Jungman:1996df,Bednyakov:1994qa}).
       Furthermore, for rather light LSP the light targets
       are preferable kinematically.
       Therefore, to search for very light dark matter
       LSP, which are not yet completely excluded 
\cite{Bednyakov:1997ax,Bednyakov:1998is}, 
      one have to use a light target nucleus
      and expect the SD interaction to dominate
\cite{Bednyakov:2000he,Bednyakov:2004xq,Bednyakov:2005qd}.

\begin{figure}[!h] %%%%%%%%%%%%%%%%%%%%%%%%%%%%%%%%%%%%%
\begin{picture}(100,80)
\put(0,-8){\includegraphics{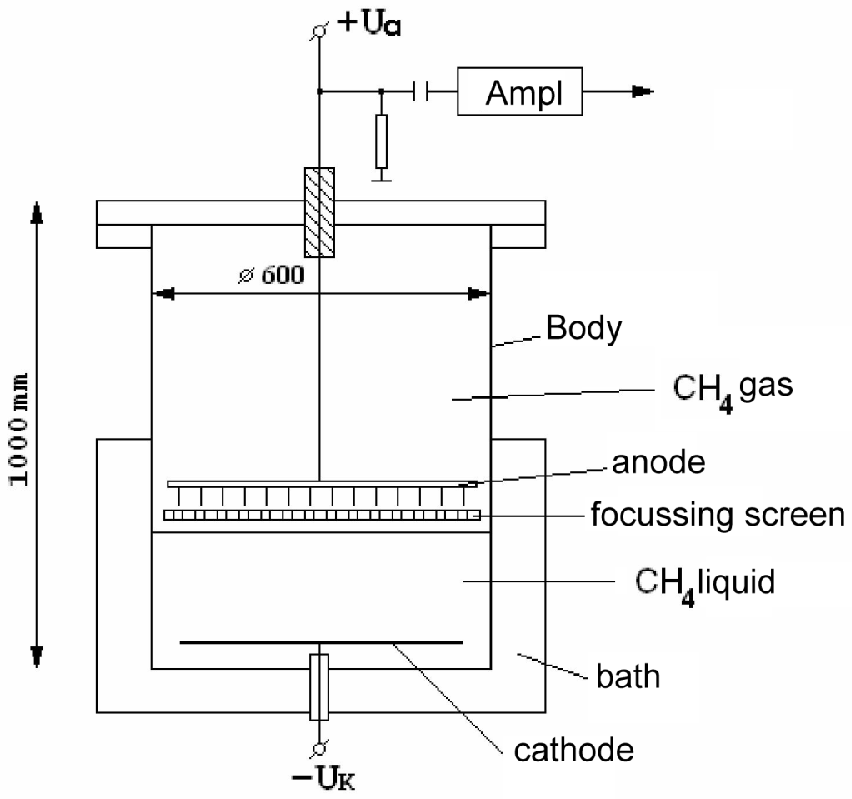}}
\end{picture}
\caption{Schematic view of the proposed two-phase methane chamber.
\label{CH4-fig1}}
%\end{figure}
%\begin{figure}[h] %%%%%%%%%%%%%%%%%%%%%%%%%%%%%%%%%%%%%
\begin{picture}(100,50)
\put(-10,-8){\includegraphics{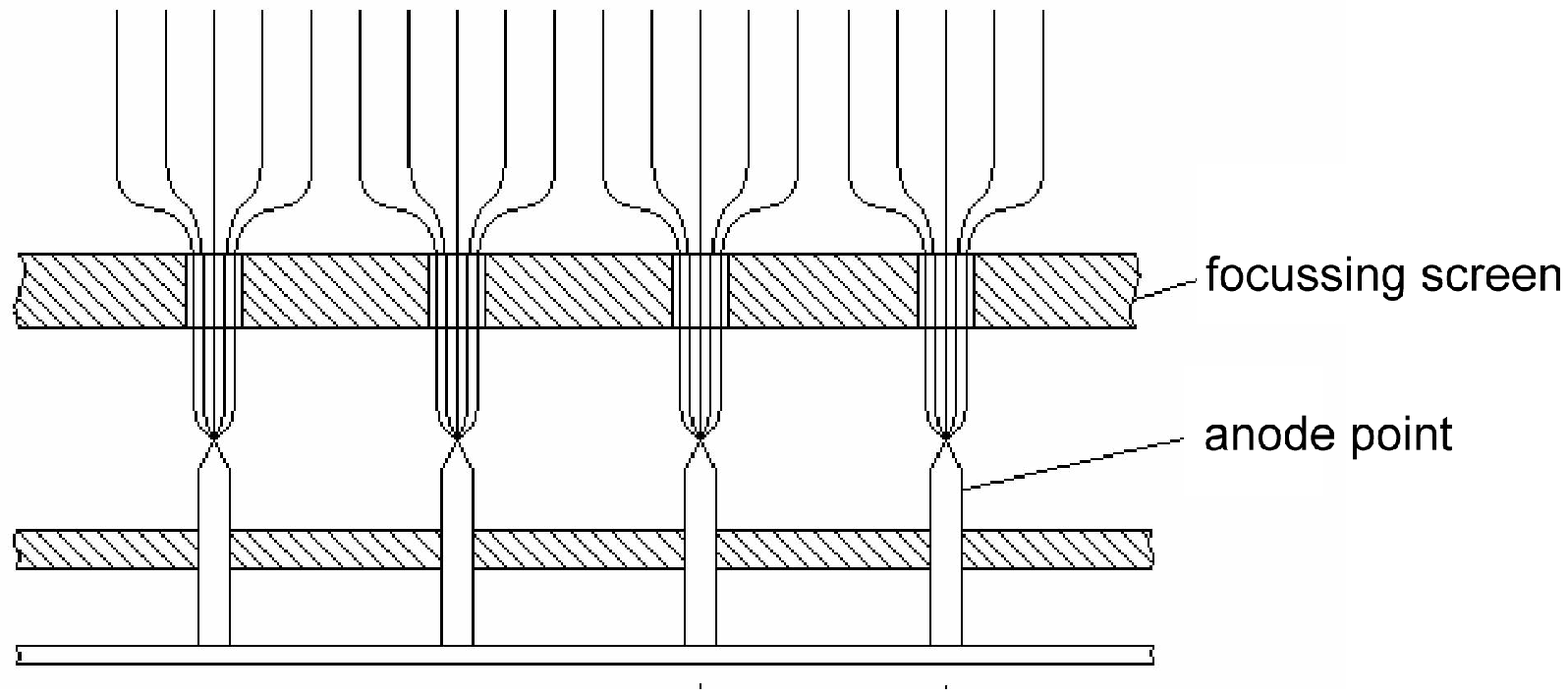}}
\end{picture}
\caption{System of focussing screen and anode points.
\label{CH4-fig2}}
\end{figure}

\enlargethispage{\baselineskip}
      In this short note 
      we describe a new special liquid-methane chamber
\cite{Barbash:1980me,Barabash:1981vh,Barabash:1982mv,Barabash:1981aa} %% [1-5].
      which could be used to search for the DM particles, 
      provided they are the lightest neutralinos. 
     The proposed two-phase methane ionization chamber is shown in 
Fig.~\ref{CH4-fig1}.
 %%%%%%%%%%%%%%%%%%%%
      The {\tt Body}\ of the chamber is made of titanium. 
      The {\tt Cathode}\ of the chamber is immersed in liquid methane. 
      The layer of liquid methane above the cathode is equal to 200 mm. 
      The temperature of liquid methane is equal to 115~K and the 
      pressure of gaseous methane over the liquid methane is equal 
      to 1.3 bar for this temperature. 
      The {\tt Anode}\  of the chamber is placed in gaseous methane 
      above liquid methane. 
      The anode consists of a system of points. 
      The {\tt Focussing screen}\ is placed between the Anode 
      and liquid methane. 
      The screen has a system of holes 
\cite{Ovchinnikov:1999nw} 
      concentrical on the relevant anode points 
(Fig.~\ref{CH4-fig2}). 
       The values of the electrical potentials on the electrodes of the 
       chamber are determined in such a way that all electrical 
       lines of force are focused on the anode points.
\begin{figure}[h!] %%%%%%%%%%%%%%%%%%%%%%%%%%%%%%%%%%%%%
\begin{picture}(100,80)
\put(-10,-5){\includegraphics{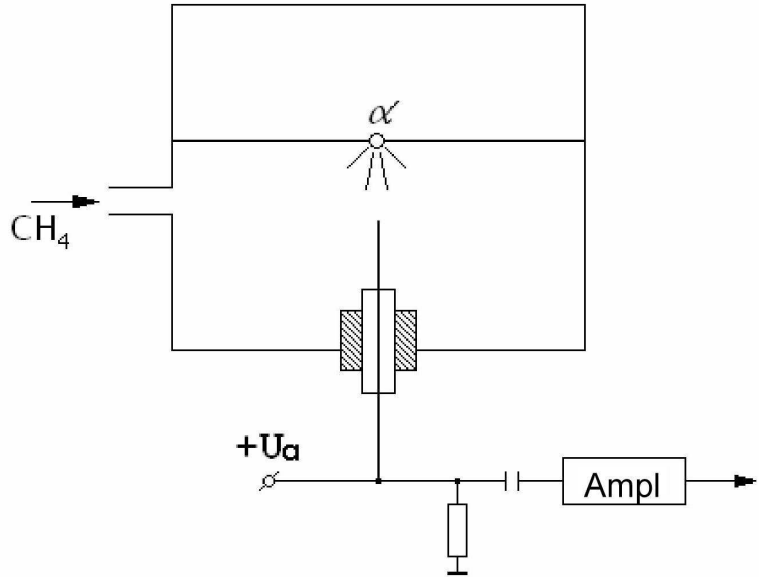}}
\end{picture}
\caption{Shematic view of the detector.
\label{CH4-fig3}
}
\end{figure} %%%%%%%%%%%%%%%%%%%%

        It is necessary to point out that the method with point 
	anodes and a focussing screen can be used in a liquid ionization 
	chamber with the Anode and the screen placed immediately 
	in the liquid medium. 
	The proportional discharge was obtained in liquid argon 
	(K$_{\rm ampl}$=100) with the point anode $\sim 0.5\mu$m in 
	diameter at the end of the Anode in work 
\cite{Kim:2004bx}.
       A model of the detector was constructed for testing 
       the work of the methane chamber with the anode points 
(Fig.~\ref{CH4-fig3}). 
       The W-wire with a diameter of 20~$\mu$m is used as the point anode 
       in this detector. 
       The Pu$^{239}$ $\alpha$-source ($10^{3}~\alpha/$s)
       is placed at the cathode. 
       The electric field intensity varies as 
       $R^{-2}$ near the point anode, 
       whereas in the cylindrical chamber it varies as $R^{-1}$. 
       This allows a greater electric field intensity 
       near point anode in semispherical geometry 
       than in the cylindrical chamber. 
       The maximum amplification factors 
       obtained in the detector with different gases are shown in 
Table~\ref{CH4-tab}. %1.
\begin{table}[h!] %%%%%%%%%%%%%%%%%%%%%%%%%%%%%%%%%%%%%%%%%%%%%%%%%
\caption{Amplification factors for different fillings of the detector.}
\label{CH4-tab}
\begin{tabular}{|c|c|c|c|c|c|}\hline
\raisebox{-1.50ex}[0cm][0cm]{Gas}& 
\multicolumn{5}{|c|}{Pressure, kgf/cm$^{2}$}  \\
\cline{2-6} 
       &   0&  5&  10&   25& 50 \\ \hline
H$_{2}$& 600&   & 600&  300&    \\ \hline
~~~~H$_{2}+10\%$CH$_{4}$~~~~& 600& & & 600& 300 \\\hline
CH$_{4}$&~~~4000~~~&~~~4000~~~& &~~~1000~~~&~~~600~~~ \\ \hline
$n$-C$_{4}$H$_{10}$& & 4000& & &  \\\hline
Ar$+10\%$CH$_{4}$ & & &~~~4000~~~& 2000&  \\\hline
\end{tabular}
\end{table} %%%%%%%%%%%%%%%%%%%%%%%%%%%%%%%%%%%%%%%%%%%%%%%%%%%%%%%%

       The mean calculated energy of recoil hydrogen atoms after 
       neutralino scattering is equal to $ \sim $1 keV, 
       of which $ \sim $80{\%} (0.8 keV) are spent for ionization 
\cite{ICRU:1993}.
       One recoil hydrogen atom of 1 keV 
       produces $\sim 20$ ionization electrons 
\cite{Verbinski:1974}.
       The amplitude of the signal at the point anode will 
       be equal to $\sim 10^{5}$ electrons 
       with the amplification factor $\sim 4\cdot 10^{3}$ 
       obtained in the model detector. 
       This signal can be easily detected.

       CD$_{4}$ can be used for filling the chamber 
       instead of CH$_{4}$ 
\cite{Barabash:1982aa}.  % [11]. 
       In principle, {CD}$_{4}$ has the following advantages over {CH}$_{4}$.
       First, the energy of the recoil deuterium atom 
       after neutralino scattering is some 2 times larger 
       than for hydrogen. %by a factor of $\sim $2.
       Secondly, the deuterium nucleus spin is  
       $J=1$ ($J=\frac{1}{2}$ for hydrogen nucleus). 
       This could increase in general 
       the spin-dependent cross section. 

\smallskip
       In conclusion,
       we give only an idea of using a specially designed 
       liquid-methane ionization chamber in an experiment
       aimed at searching for the (mostly) low-mass dark-matter particles 
       on the basis of their spin-dependent interactions with hydrogen. 
       Further sumulations of all possible backgrounds 
       are obviously necessary to make a conclusion as to real 
       prospects of such a full-scale dark-matter experiment with 
       liquid-methane ionization chamber.
 
\smallskip
        One of the authors (V.B.) thanks the RFBR (02--02--04009) 
	for support.

\input{CH4-DM.bbl}

%%%%%%%%%%%%%%
\end{document}

%% file: CH4-DM.bbl
\providecommand{\href}[2]{#2}\begingroup\raggedright\endgroup